# Comparing single-node and multi-node performance of an important fusion HPC code benchmark

Emily A. Belli

General Atomics, La Jolla, CA, USA, bellie@fusion.gat.com

Jeff Candy

General Atomics, La Jolla, CA, USA, candy@fusion.gat.com

Igor Sfiligoi

University of California San Diego, La Jolla, CA, USA, isfiligoi@sdsc.edu

Frank Würthwein

University of California San Diego, La Jolla, CA, USA, fkw@ucsd.edu

Fusion simulations have traditionally required the use of leadership scale High Performance Computing (HPC) resources in order to produce advances in physics. The impressive improvements in compute and memory capacity of many-GPU compute nodes are now allowing for some problems that once required a multi-node setup to be also solvable on a single node. When possible, the increased interconnect bandwidth can result in order of magnitude higher science throughput, especially for communication-heavy applications. In this paper we analyze the performance of the fusion simulation tool CGYRO, an Eulerian gyrokinetic turbulence solver designed and optimized for collisional, electromagnetic, multiscale simulation, which is widely used in the fusion research community. Due to the nature of the problem, the application has to work on a large multi-dimensional computational mesh as a whole, requiring frequent exchange of large amounts of data between the compute processes. In particular, we show that the average-scale nl03 benchmark CGYRO simulation can be run at an acceptable speed on a single Google Cloud instance with 16 A100 GPUs, outperforming 8 NERSC Perlmutter Phase1 nodes, 16 ORNL Summit nodes and 256 NERSC Cori nodes. Moving from a multi-node to a single-node GPU setup we get comparable simulation times using less than half the number of GPUs. Larger benchmark problems, however, still require a multi-node HPC setup due to GPU memory capacity needs, since at the time of writing no vendor offers nodes with a sufficient GPU memory setup. The upcoming external NVSWITCH does however promise to deliver an almost equivalent solution for up to 256 NVIDIA GPUs.

CCS CONCEPTS •Computer systems organization~Architectures~Distributed architectures •Networks~Network performance evaluation •Applied computing~Physical sciences and engineering

**Additional Keywords and Phrases:** High Performance Computing, interconnect bandwidth, benchmarking, Fusion science

## 1 INTRODUCTION

The advent of compute accelerators, namely server-class Graphics Compute Units (GPUs), has provided drastic improvements in compute density for modern High Performance Computing (HPC) systems. Each node of the system now provides over an order of magnitude higher compute throughput than the CPU-only nodes they replaced. Modern GPUs also implement fast interconnects, allowing for effective use of multiple GPUs for tightly coupled applications.

Unfortunately, the off-node interconnect capacities, i.e. network adapters, did not improve at the same rate, drastically reducing the effectiveness of multi-node setups for communication-heavy scientific applications.

To showcase this discrepancy, we provide both single-node and multi-node benchmark results for the fusion simulation tool CGYRO[1,2], an Eulerian gyrokinetic turbulence solver designed and optimized for collisional, electromagnetic, multiscale simulation, which is widely used in the fusion research community. Due to the nature of the problem, the application has to work on a large multi-dimensional computational mesh as a whole, requiring frequent exchange of large amounts of data between the compute processes. High interconnect throughput between compute resources is thus as important as the compute speed of those resources. The average-scale nl03 benchmark CGYRO simulation can now be computed in reasonable time on a single node equipped with 16 NVIDIA A100 40GB GPUs, which is available as a virtual instance on the Google Cloud (GCP) [3]. We then describe multi-node setups that provide comparable time-to-solution at leadership-class HPC systems operated by the Oak Ridge National Laboratory (ORNL) and the National Energy Research Scientific Computing Center (NERSC), emphasizing the drastic reduction in compute resource effectiveness when moving from a single-node to a multi-node setup. It should be noted that those HPC systems do not offer compute nodes that could compute the nl03 benchmark case on a single node in reasonable time.

We acknowledge that while a large fraction of CGYRO simulations can fit in the tested GCP instance, the most cutting-edge simulations cannot and will continue to require multi-node, leadership-class HPC systems. That said, there are currently no on-prem resources that provide in a single node the amount of compute and GPU memory the GCP instance does. The aim of this paper is to encourage resource providers serving the fusion community to start offering compute resources of this kind, and beyond, as the drastic increase in resource effectiveness for a non-negligible subset of problems benefits all parties.

## 2 THE FUSION SIMULATION TOOL CGYRO

Fusion energy research has made significant progress over the years, yet many fundamental aspects of its physics are still not well understood. While experimental methods are essential for gathering new operational modes, simulations are used to validate basic theory, plan experiments, interpret results on present devices, and ultimately to design future devices.

The CGYRO simulation tool operates on a 6-dimensional grid (3D space + 2 D velocity + 1 D species). The numerical discretization is spectral in the radial and binormal dimensions ($k_x$, $k_y$), pseudospectral in both velocity dimensions, and uses a unique 5th-order conservative upwind scheme in θ, the coordinate along the field lines. In order to concurrently utilize multiple compute resources, the problem is split in many smaller sub-problems using the Message Passing Interface (MPI) paradigm, splitting the grid using two orthogonal MPI communicators [4]. The first MPI communicator size is fixed by the problem size; it is always the value of N_TOROIDAL (i.e. $k_y$ /2). All the other dimensions are then lumped together and can be distributed along the other MPI communicator. At each time step, there are several interleaved MPI_AllToAll and MPI_AllReduce collective operations needed on the two MPI communicators. The MPI_AllToAll on the first communicator is typically the most data intensive, although the MPI_AllReduce data volume grows with the number of MPI processes and would eventually become dominant.

The ability of splitting the problem in very small sub-problems gives CGYRO a very large dynamic range, allowing it to exploit HPC systems from 10s to 1000s of independent compute units, e.g. CPUs or GPUs. Note, however, that the users are not required to assign a single MPI process per compute unit; all multi-core CPUs and larger GPUs are capable of serving several compute processes in parallel. This later mode of problem splitting has several advantages, but for the purpose of this paper we will stress the resulting reduction in communication traffic on the off-chip interconnects. Since the amount of data exchanged in the two orthogonal MPI communicators is significantly different and on-chip MPI communication is significantly cheaper than communication over the external interconnect, grouping several MPI processes belonging to the first MPI communicator on the same compute resource effectively reduces the total amount of traffic on the slower interconnects in most cases. But even with this optimization, the amount of data exchanged is substantial, as we will showcase later in the paper.

Fusion scientists use CGYRO for simulating a wide range of plasma phenomena, and it would be impractical to provide benchmark results for all of them. We thus focus on the nl03 benchmark input [5], which is both representative of the mainstream CGYRO simulation at the time of writing and tightly fits in the tested single-node GCP instance. Note that the main limiting factor for CGYRO simulation is the amount of available memory, and in particular GPU memory for GPU-based systems, as the (decomposed) collision matrix must fit in it, so it is a hard requirement.

## 3 SINGLE NODE PERFORMANCE

In order to achieve a reasonable runtime we considered only GPU-based systems for this section. Moreover, the nl03 benchmark case requires at least 600 GB of GPU memory, and the best match for these requirements was the GCP a2-megagpu-16g instance [6]. It is composed of 16 NVIDIA A100-SMX4-40GB GPUs, each with a 2.4 Tbps NVLINK interconnect, which are connected through a NVSWITCH setup. The node also sports two Intel Xeon Skylake CPUs, which are connected to the GPUs using a set of 125 Gbps PCIe Gen3 links. The node is also equipped with a 100Gbps network interface, but we did not use it during the benchmark tests.

We compiled CGYRO locally using the NVIDIA SKD Toolkit and also used the included OpenMPI v3 libraries for inter-process communication. Since this was a single-node setup, no additional resource manager was needed.

Using this setup, the simulation was progressing at about 520 ms per step, out of which about 300 ms, or 57% was spent doing actual compute, and the rest was spent in inter-process communication. As can be seen, even on a single node setup the amount of time spent in communication is non-negligible, which is not entirely surprising given that during each step each GPU exchanges about 6.8GB of data with the others, mostly using the MPI_AllToAll collective operations.

## 4 MULTI-NODE HPC PERFORMANCE

We also have access to several leadership-class HPC systems, where we can run the nl03 benchmark CGYRO simulation. We chose to benchmark three of the most recent systems, namely the A100 GPU-based NERSC Perlmutter [6], the V100 GPU-based ORNL Summit [7] and the Xeon Phi CPU-based NERSC Cori [8]. We include a CPU-based system to emphasize the importance of keeping the interconnect and compute relative performance well balanced.

NERSC Perlmutter Phase 1 nodes each contain 4 NVIDIA A100-SXM4-40GB GPUs, each with three 800 Gbps NVLINK interconnects in a full mesh configuration. Each node also sports one AMD EPYC Milan CPU, which is connected to the GPUs using a set of 250 Gbps PCIe Gen4 links. The network interconnect of each node is provided by two 100 Gbps HPE Cray Slingshot interfaces. The system contains over 1,000 nodes, but we provide benchmark data for only up to 16 of them.

ORNL Summit nodes each contain 6 NVIDIA V100-SMX2-16GB GPUs and 2 IBM POWER9 CPUs. Each CPU is connected with 3 GPUs using a set of 400 Gbps NVLINK inter-connects in a full mesh configuration. The two NVLINK domains are then linked with a 512 Gbps interconnect inside the node. The network interconnect of each node is provided by two 100 Gbps EDR InfiniBand interfaces. The system contains over 4,000 nodes, but we provide benchmark data for only up to 32 of them.

NERSC Cori is an Intel Xeon Phi, also known as Knight Landing (KNL), CPU-based system with 96 GB of memory per node. Each node sports a single CPU and a 40 Gbps Aries interconnect link. The whole system is composed of over 9,000 nodes, but we provide benchmark data for only up to 256 of them.

We compiled CGYRO on each system using the system-provided tools. The NVIDIA HPC Toolkit compiler was used on the two GPU-based system, and the Intel Compiler was used on the CPU-based system. The default, recommended pre-installed MPI library and the default batch system were used on each of the three systems.

Due to insufficient GPU memory capacity, none of the nodes in these tested GPU-based HPC systems is big enough to run the nl03 benchmark CGYRO simulation on its own, so we had to run in a multi-node setup. And due to order-of-magnitude difference in compute performance between CPUs and GPUs we did not even consider CPU-only single nodes. On the other hand, the simulation is small enough to only use a tiny fraction of the nodes in any of the HPC systems, so we do not have to worry about the effects of the network topology.

For each HPC system we picked two node counts that provided a simulation time per CGYRO step comparable to the tested single-node setup described above. In particular, we used 8 and 16 nodes on Perlmutter, 16 and 32 nodes on Summit and 128 and 256 nodes on Cori. We also added an additional Perlmutter data point where we used only two of the four GPUs, to showcase the impact of different deployment options. The measured runtimes of the three HPC systems, as well as of the GCP instance, are available in Table 1. The same table also contains the amount of data exchanged during each simulation step by each GPU and each node, so one can correlate that to the measured time spent in inter-process communication. As mentioned in Section 2, the amount of data exchanged at each step changes as one varies the level of MPI parallelism.

Table 1: Average per nl03 CGYRO simulation step runtimes, as measured on the four systems, alongside the amount of data exchanged by each GPU and each node at each simulation step. The chosen multi-node data points all have an average step time that is comparable to the single-node one.

| System | Number of nodes | Number of GPUs | Average step time | Average compute time | Average comm. time | Step data per GPU | Step data per node | Fraction in compute |
|---|---|---|---|---|---|---|---|---|
| GCP a2-megagpu-16g | 1 | 16 | 520 ms | 300 ms | 170 ms | 6.8 GB | N/A | 57% |
| Perlmutter Phase 1 | 8 | 32 | 670 ms | 170 ms | 470 ms | 3.7 GB | 9.5 GB | 25% |
| Perlmutter Phase 1 | 16 | 64 | 360 ms | 81 ms | 250 ms | 2.0 GB | 5.2 GB | 23% |
| Perlmutter Phase 1 Using only 2xA100 | 16 | 32 | 460 ms | 160 ms | 270 ms | 3.7 GB | 6.6 GB | 35% |
| Summit | 16 | 96 | 540 ms | 100 ms | 410 ms | 1.4 GB | 5.9 GB | 19% |
| Summit | 32 | 192 | 370 ms | 54 ms | 280 ms | 0.7 GB | 3.2 GB | 15% |
| Cori | 128 | N/A | 750 ms | 410 ms | 300 ms | N/A | 1.2 GB | 55% |
| Cori | 256 | N/A | 530 ms | 250 ms | 220 ms | N/A | 0.7 GB | 47% |

A graphic representation of the same data is also available in Figure 1, from which it is obvious that moving from single-node to multi-node GPU setups incurs a significant inter-process communication penalty, i.e. the ratio between time spent on compute vs total time spent for each step (labeled as "Fraction in compute") is significantly lower for the multi-node setups. This is especially obvious when each node hosts many GPUs, which have a significantly lower ratio between compute throughput and network bandwidth, and thus result in a much lower ratio between time spent on compute vs total simulation time. The CGYRO simulation progress thus becomes completely network bandwidth bound.

It is also interesting to note that the simulation run on the single GCP node setup spends about the same fraction of time on compute as the multi-node CPU-based Cori system. The 10x compute throughput difference between a single GPU and a single CPU results in a comparable increase in inter-process data volume, saturating even the node-local NVLINK interconnect.

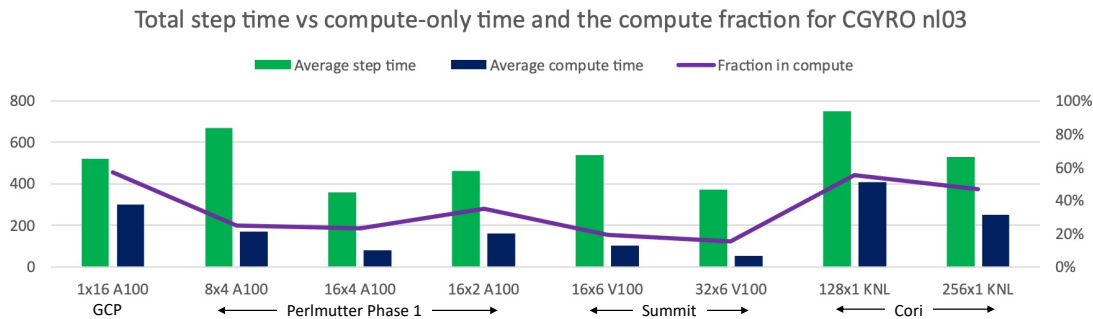

Figure 1: Graphic representation of data from Figure 1, with an emphasis on fraction of time spent on compute-only.

## 5 DISCUSSION AND CONCLUSIONS

The switch from CPU-based to GPU-based HPC systems has provided a great improvement in computational throughput for many scientific applications. The increased computational throughput has however not been matched with the equivalent increase in interconnect throughput, especially networking on multi-node setups. While Cori had 40 Gbps of network bandwidth per CPU, Summit has less than 35 Gbps per GPU, even though each Summit's GPU is more than 5x faster that each Cori's CPU. The situation is slightly better on Perlmutter Phase 1 nodes, with 50 Gbps per GPU, although Perlmutter's GPUs are slightly faster than those on Summit. The net result is only limited speedup for the benchmarked fusion simulation tool CGYRO, whose algorithm is communication-heavy.

    The increased computational power of GPUs does however imply that a single node, equipped with many GPUs, can now deliver the same simulation throughput as about ten (smaller) GPU-based HPC nodes, or hundreds of CPU-only HPC nodes. Since node-local, GPU-to-GPU interconnects are much faster than node-to-node interconnects, CGYRO problems that can fit in a single node can still fully benefit from the computational throughput improvements, resulting in comparable simulation times using less than half the number of GPUs. We verify this assertion by benchmarking the nl03 test case, which is representative of the mainstream CGYRO fusion simulations at the time of writing.

    Finding nodes large enough to fit the desired problem is however not trivial. Recent HPC systems seem to prefer to deploy a larger number of smaller nodes, and we had to rent an instance from a commercial Cloud provider to run the nl03 benchmark case, instead. While the path chosen by the leadership-scale HPC centers is probably a good one for solving large, cutting-edge HPC problems, since most of the benefits of a large node are lost when one has to use many of them, we believe it is a missed opportunity for effectively solving mainstream fusion problems, which represent the bulk of CGYRO compute by number.

    In summary, we show that large GPU-based nodes provide excellent value for mainstream fusion CGYRO simulations. And we expect that other communication-heavy HPC applications working on modest-sized problems would behave similarly. We thus encourage resource providers to consider such setups when procuring new systems, especially when supporting fusion science compute.

Looking forward, we would also like to see systems that support CGYRO simulations that are just slightly too big for a single node, too, i.e. fusion simulations that may need up to a handful of large nodes. Currently the best option is to attach one high-end network card to each of the GPUs, like it is being done by the NVIDIA DGX [9] systems; while more expensive and less performant that a single-node setup, it would likely still be the most effective multi-node setup for such simulations. Looking slightly further at the future, we are excited by NVIDIA's external NVSWITCH [10] announcement, with its promise of extending node-local interconnect speeds to up to 32 nodes, which would be ideal for solving these other fusion problems. As before, we expect other communication-heavy HPC applications to benefit as well.

**ACKNOWLEDGMENTS**

This work was partially supported by the U.S. Department of Energy under awards DE-FG02-95ER54309, DE-FC02-06ER54873 (Edge Simulation Laboratory) and DE-SC0017992 (AToM SciDAC-4 project), and the US National Science Foundation (NSF) Grant OAC-1826967. An award of computer time was provided by the INCITE program. This research used resources of the Oak Ridge Leadership Computing Facility, which is an Office of Science User Facility supported under Contract DE-AC05-00OR22725. Computing resources were also provided by the National Energy Research Scientific Computing Center, which is an Office of Science User Facility supported under Contract DE-AC02-05CH11231.